\documentclass[12pt]{article}
\usepackage{amssymb}
\usepackage{graphicx}
\def\be{\begin{equation}}
\def\ee{\end{equation}}

\begin{document}
\begin{center}
\hfill \vbox{
\hbox{April 2007}}
\vskip 0.8cm
{\Large\bf Screening masses in the SU(3) pure gauge theory and universality}\\ 
\end{center}
\vskip 0.3cm
\centerline{R. Falcone$^*$, R. Fiore$^*$, M. Gravina$^*$ and
A. Papa$^*$}
\vskip 0.3cm
\centerline{\sl Dipartimento di Fisica, Universit\`a della Calabria,}
\centerline{\sl and Istituto Nazionale di Fisica Nucleare, Gruppo collegato
di Cosenza}
\centerline{\sl I--87036 Arcavacata di Rende, Cosenza, Italy}
\vskip 0.6cm
\begin{abstract}
We determine from Polyakov loop correlators the screening masses in the deconfined 
phase of the (3+1)$d$ SU(3) pure gauge theory at finite temperature near the transition, 
for two different channels of angular momentum and parity. 
Their ratio is compared with that of the massive excitations with the same quantum numbers
in the 3$d$ 3-state Potts model in the broken phase near the transition point at zero 
magnetic field. 
Moreover we study the inverse decay length of the correlation between the real parts and 
between the imaginary parts of the Polyakov loop and compare the results with expectations
from perturbation theory and mean-field Polyakov loop models.
\end{abstract}

\vfill
\hrule
\vspace{0.3cm}
\noindent$^*$ {\it e-mail addresses}: 
rfalcone,fiore,gravina,papa@cs.infn.it 

\newpage

\section{Introduction}

The essential features of confinement, of the deconfinement phase transition at finite 
temperature and of the deconfined phase in QCD can be studied without loss of 
relevant dynamics in the regime of infinitely massive quarks or, equivalently, 
in the SU(3) {\em pure} gauge theory.

The (3+1)$d$ SU(3) pure gauge theory at finite temperature undergoes a 
confinement/deconfinement phase transition associated with the 
breaking of the center of the gauge group, Z(3)~\cite{Polyakov:vu,Susskind:up}. 
This transition is {\em weakly} first 
order~\cite{Fukugita:1985,Bacilieri:1988,Brown:1988,Kogut:1983,
Kaczmarek} and the order parameter is the Polyakov loop~\cite{Polyakov:vu,Susskind:up}.

In general, the deconfinement transition in the $(d+1)$-dimensional SU(N) pure gauge theory at 
finite temperature can be put in relation with the order/disorder~\cite{Gavai-Karsch-Petersson} 
phase transition of the $d$-dimensional Z(N) spin model through the Svetitsky-Yaffe 
conjecture~\cite{Svetitsky:1982gs}.
According to this conjecture, when the transition is {\it second order} the gauge theory and 
the spin model belong to the same universality class and therefore share critical indices, 
amplitude ratios and correlation functions at criticality. 
There are so many numerical evidences in favor of this conjecture -- see, for instance, 
Refs.~\cite{Gliozzi:1997yc,Fiore:1998uk,Engels:1998nv,Fortunato:2000hg,
Fiore:2001ci,Fiore:2001pf,Papa:2002gt} and, for a review, Ref.~\cite{Pelissetto-Vicari:2002} --
that there is no residual doubt about its validity. 

In the case of {\it first order} phase transitions, as for the (3+1)$d$ SU(3) pure gauge
theory, the Svetitsky-Yaffe conjecture is not applicable in strict sense, since the correlation 
length keeps finite at the transition point and a universal behavior, independent of the details 
of the microscopic interactions, cannot be established. Therefore there are no {\it a priori} 
reasons why the critical dynamics of (3+1)$d$ SU(3) and of its would-be counterpart spin model, 
3$d$ Z(3) or 3-state Potts model~\cite{Blote-Swendsen,Janke-Villanova}, should look similar. 
It turns out, however, that for both these theories the phase transition is {\it weakly} first 
order, which means that the correlation length at the transition point, although finite, 
becomes much larger than the lattice spacing. This leads to the expectation that the two 
theories may have some long distance aspects in common and 
that some specific issues of SU(3) in the transition region can be studied taking the 
3$d$ 3-state Potts model as an effective model.

Since there are no critical indices to compare, a test of this possibility could be based
on the comparison of the spectrum of massive excitations of the 3$d$ 3-state Potts model 
in the broken phase near the transition at zero magnetic field with the spectrum of the inverse 
decay lengths of Polyakov loop correlators in the (3+1)$d$ SU(3) gauge theory in the deconfined phase 
near the transition. If universality would apply in strict sense, these spectra should exhibit 
the same pattern, as suggested by several numerical determinations in the 3$d$ 
Ising class~\cite{Caselle:1999tm,Caselle:2001im,Fiore:2002fj,Fiore-Papa-Provero-2003}. 

In this paper we determine the low-lying masses of the spectrum of the (3+1)$d$ SU(3) pure gauge 
theory at finite temperature in the deconfined phase near the transition in two different sectors 
of parity and orbital angular momentum, 0$^+$ and 2$^+$, and compare their ratio with that of 
the corresponding massive excitations in the phase of broken Z(3) symmetry of the 
3$d$ 3-state Potts model, determined in Ref.~\cite{FFGP06}.

In fact, we extend our numerical analysis to temperatures far away from the transition
temperature $T_t$ in order to look for possible ``scaling'' of the fundamental masses
with temperature.
Moreover, we consider also the screening masses resulting from correlators of the real parts
and of the imaginary parts of the Polyakov loop. These determinations can represent useful
benchmarks for effective models of the high-temperature phase of SU(3), such as those based
on mean-field theories of the Polyakov loop, suggested by R.~Pisarski~\cite{Pisarski0101168}.

The paper is organized as follows: in Section~\ref{sec:masses} we describe the methods used to 
extract the screening masses through Polyakov loop correlators in (3+1)$d$ SU(3); 
in Section~\ref{results} we present our numerical results and compare them with expectations
from the 3$d$ 3-state Potts model and from Polyakov loop effective models; in the last 
Section we draw our conclusions.

\section{Screening masses from Polyakov loop correlators}
\label{sec:masses}

In this Section we define the general procedure for extracting screening masses 
in (3+1)$d$ SU(3) at finite temperature from correlators of the Polyakov loop, 
\begin{equation}
P(x,y,z) =\mbox{Tr}\prod_{n_t =1}^{N_t}U_4(x,y,z,n_t a)\ ,
\end{equation}
where $N_t$ is the number of lattice sizes in the temporal direction and $a$ is the lattice spacing.

The point-point connected correlation function is defined as
\begin{equation}
\Gamma_{i_0}(r)=\langle P_i^\dagger P_{i_0} \rangle -\langle P_i^\dagger \rangle 
\langle P_{i_0} \rangle \ ,
\end{equation}
where $i$ and $i_0$ are the indices of sites and $r$ is the distance between them. 
The large-$r$ behavior of the point-point connected correlation function is determined by
the correlation length of the theory, $\xi_0$, or, equivalently, by its inverse, the 
fundamental mass. In order to extract the fundamental mass it is convenient to define the 
wall-wall connected correlation function, since numerical data in this case can be directly 
compared with exponentials in $r$, without any power prefactor. 

The 0$^+$-channel connected wall-wall correlator in the $z$-direction is defined as
\begin{equation}
G(|z_1-z_2|)=\mbox{Re}\langle \bar{P}(z_1) \bar{P}(z_2)^\dagger \rangle -|\langle P \rangle|^2\ ,
\label{corr0}
\end{equation}
where
\begin{equation}
\bar{P}(z)=\frac{1}{N_xN_y}\sum_{n_x=1}^{N_x} \sum_{n_y=1}^{N_y}P(n_x a,n_y a,z)\ ,
\label{wall0}
\end{equation} 
represents the Polyakov loop averaged over the $xy$-plane at a given $z$. Here and in the following,
$N_i$ ($i=x, y, z$) is the number of lattice sites in the $i$-direction.
The wall average implies the projection at zero momentum in the $xy$-plane.

The correlation function~(\ref{corr0}) takes contribution from all the screening
masses in the 0$^+$ channel. In fact, the general large distance behavior for the function 
$G(|z_1-z_2|)$, in an infinite lattice, is: 
\begin{equation}
G(|z_1-z_2|)= \sum_n a_n e^{-m_n |z_1-z_2|}\ ,
\label{corr-funct}
\end{equation}
where $m_0$ is the fundamental mass, while $m_1$, $m_2$, ... are higher masses
with the same angular momentum and parity (0$^+$) quantum numbers of the fundamental mass.
On a periodic lattice the above equation must be modified by the inclusion of the so called
``echo'' term:
\begin{equation}
G(|z_1-z_2|)= \sum_n a_n\biggl[e^{-m_n |z_1-z_2|}+e^{-m_n (N_z-|z_1-z_2|)}\biggr]\;.
\label{corr-funct_echo}
\end{equation}
The fundamental mass in a definite channel can be extracted from wall-wall 
correlators by looking for a plateau of the effective mass,
\begin{equation}
m_{\mbox{\footnotesize eff}}(z)= \ln \frac{G(z-1)}{G(z)}
\label{m_eff}
\end{equation}
at large distances.
For the 2$^+$-channel, we used the variational method~\cite{Kronfeld,Luscher-Wolff}, which 
consists in defining several wall-averaged operators with 2$^+$ quantum numbers and in building 
the matrix of cross-correlations between them. The eigenvalues of this matrix are 
single exponentials of the effective masses in the given channel. The variational
method is usually adopted to determine massive excitations above the fundamental one
in the given channel. In our case, however, it turned out that this method improved
the determination of the fundamental mass in the 2$^+$-channel, but did not lead to determine
higher excitations. Our choice of wall-averaged operators in the 2$^+$-channel is
inspired by Ref.~\cite{Caselle1997} and reads 
\begin{equation}
\bar{P}_n(z)=\frac{1}{N_xN_y}\sum_{n_x=1}^{N_x} \sum_{n_y=1}^{N_y}P(n_x a,n_y a,z)
\biggl[P(n_x a+na,n_y a,z)-P(n_x a,n_y a+na,z)\biggr]\ .
\label{wall2}
\end{equation} 
In most cases we have taken 8 operators, corresponding to different values of $n$, with the 
largest $n$ almost reaching the spatial lattice size $N_x$. 
We have determined the 0$^+$ and the 2$^+$ fundamental masses in lattice units, 
$\hat m_{0^+}$ and $\hat m_{2^+}$, over a wide interval of temperatures above the transition 
temperature $T_t$ of (3+1)$d$ SU(3) and have studied 

- if in some region, lying {\it strictly above} $T_t$, a scaling law can be found
for the fundamental correlation length $\hat\xi_0=1/\hat m_{0^+}$;

- if $\hat m_{2^+}$ has the same scaling behavior as $\hat m_{0^+}$;

- how the ratio $m_{2^+}/m_{0^+}$ compares with the ratio of the corresponding 
excitations in the broken phase of 3$d$ 3-state Potts model near the transition, 
$m_{2^+}/m_{0^+}=2.43(10)$, found in Ref.~\cite{FFGP06} on a 
48$^3$ lattice.

We consider also correlators of the (wall-averaged) real and imaginary parts of the Polyakov 
loop, defined as
\begin{eqnarray}
G_R(|z_1-z_2|) &=& \langle \mbox{Re} \bar P(z_1) \mbox{Re} \bar P(z_2)\rangle - 
\langle \mbox{Re} \bar P(z_1)\rangle \langle \mbox{Re} \bar P(z_2)\rangle \;,
\label{Dumitru2:1}\\
G_I(|z_1-z_2|) &=& \langle \mbox{Im} \bar P(z_1) \mbox{Im} \bar P(z_2)\rangle\;.
\label{Dumitru2:2}
\end{eqnarray}
The corresponding screening masses, $\hat m_R$ and $\hat m_I$, can be extracted in the 
same way as for the 0$^+$ mass.

We have studied the ratio $m_I/m_R$ over a wide interval of temperatures above the 
transition temperature $T_t$ of (3+1)$d$ SU(3) and seen how it compares with the prediction from 
high-temperature perturbation theory, according to which it should be equal to 
3/2~\cite{Nadkarni:1986cz,Dumitru:2002cf}, and with the prediction from the mean-field 
Polyakov loop model of Ref.~\cite{Pisarski0110214}, according to which it should be equal to 3 in 
the transition region. The interplay between the two regimes should delimit the 
region where mean-field Polyakov loop models should be effective.

\section{Numerical results}
\label{results}

We used the Wilson lattice action and generated Monte Carlo configurations by a combination 
of the modified Metropolis algorithm~\cite{Cabibbo-Marinari} with over-relaxation on
SU(2) subgroups~\cite{Adler}. The error analysis was performed by the jackknife method over bins 
at different blocking levels. We performed our simulations on a 16$^3\times$4 lattice, for which 
$\beta_t=5.6908(2)$~\cite{Boyd:1996bx}, over an interval of $\beta$ values ranging from
5.69 to 9.0. 

\begin{figure}[tb]
\centering
\includegraphics[width=11.5cm, bb=40 40 700 620]{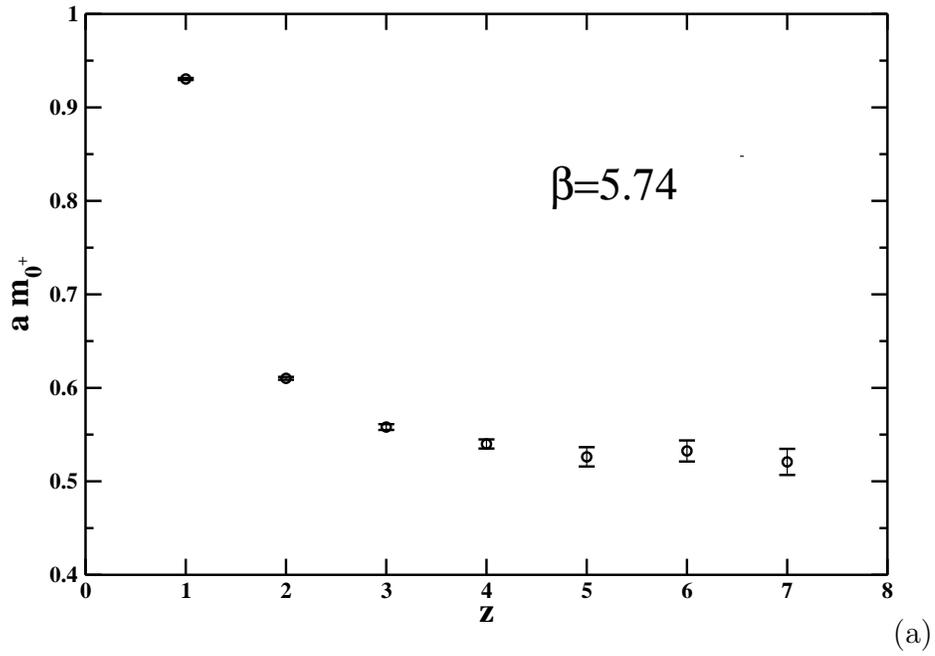} (a) 
\includegraphics[width=11.5cm, bb=40 40 700 620]{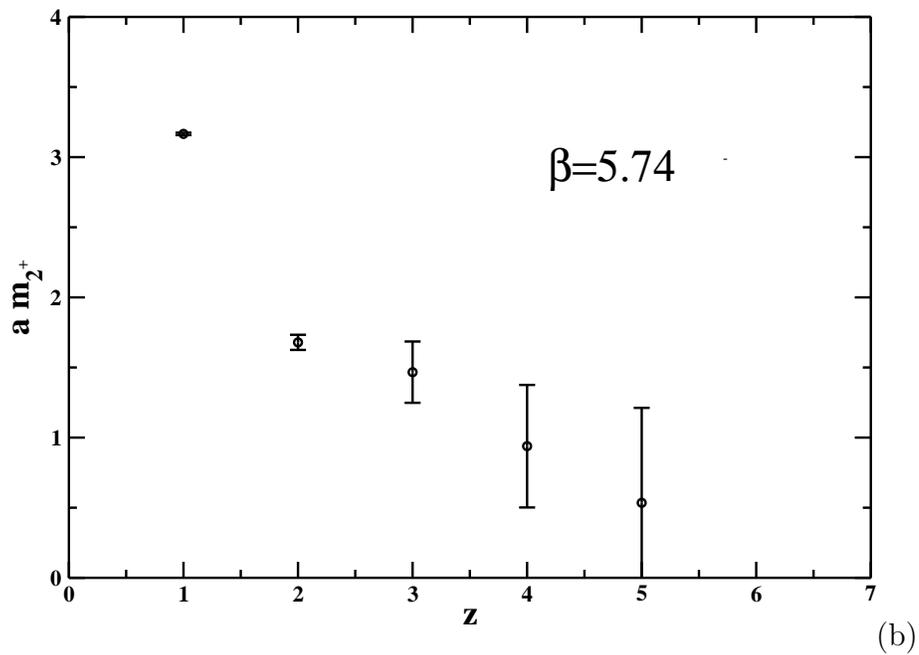} (b) 
\caption[]{Effective mass in the 0$^+$ (a) and the 2$^+$ (b) channel as a function 
of the separation between walls on the $(x,y)$ plane at $\beta=5.74$.}
\label{eff_masses_5.74} 
\end{figure}

\begin{figure}[tb]
\centering
\includegraphics[width=11.5cm, bb=40 40 700 620]{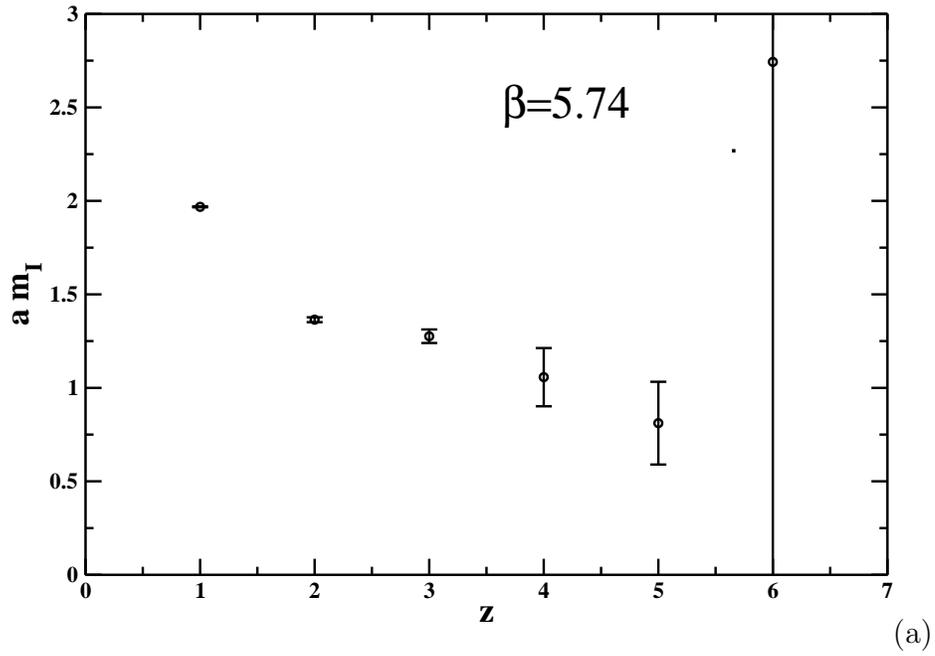} (a)
\includegraphics[width=11.5cm, bb=40 40 700 620]{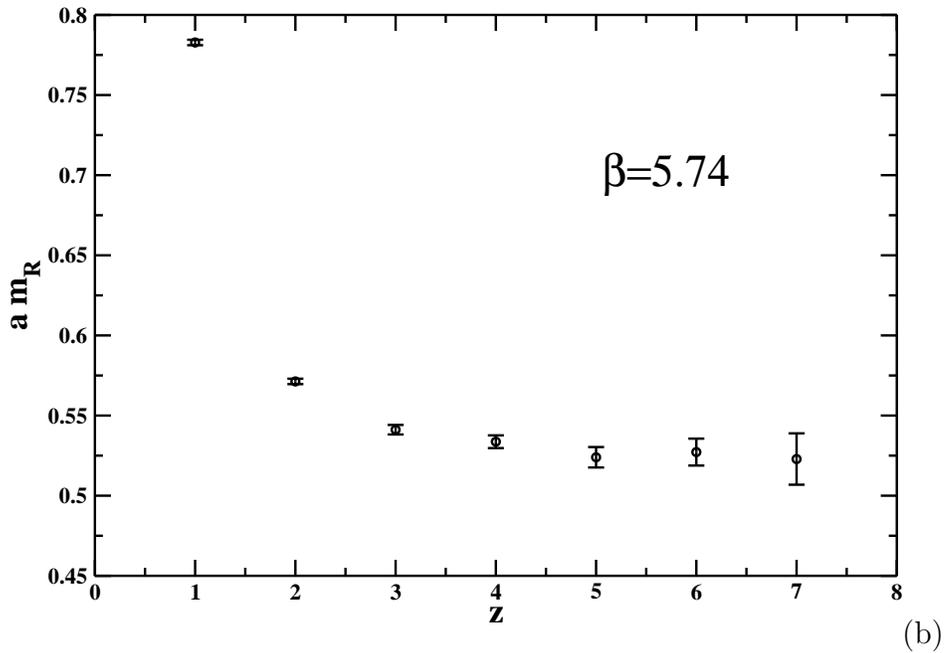} (b)
\caption[]{Effective mass $\hat m_I$ (a) and $\hat m_R$ (b) as a function of the separation between 
walls on the $(x,y)$ plane at $\beta=5.74$.}
\label{eff_masses_ri_5.74} 
\end{figure}

Screening masses are determined from the plateau of $m_{\mbox{\footnotesize eff}}(z)$ as a 
function of the wall separation $z$. In each case, the {\it plateau mass} is taken as the 
effective mass (with its error) belonging to the {\it plateau} and having the minimal 
uncertainty. We define {\it plateau} the largest set of consecutive data points, consistent with 
each other within 1$\sigma$.
This procedure is more conservative than identifying the plateau mass and its error
as the results of a fit with a constant on the effective masses $m_{\mbox{\footnotesize eff}}(z)$,
for large enough $z$. To illustrate our definition of {\it plateau mass}, let us apply it 
to Figs.~\ref{eff_masses_5.74}(b) and~\ref{eff_masses_ri_5.74}(a), which give respectively 
the $\hat m_{2^+}$ and the $\hat m_I$ effective masses at $\beta=5.74$ versus the separation $z$ 
between walls (the cases of $\hat m_{2^+}$ and $\hat m_I$ are usually the most problematic
for the determination of the plateau mass). In Fig.~\ref{eff_masses_5.74}(b), according to
our definition, the plateau is formed by the points at $z=3$, 4 and 5, since they are compatible
each other within 1$\sigma$; the plateau mass is taken as the effective mass at $z=3$,
since this effective mass has the smallest uncertainty among the three masses
belonging to the plateau. In Fig.~\ref{eff_masses_ri_5.74}(a), according to
our definition, the plateau is formed by the points at $z=4$, 5 and 6, since they are compatible
each other within 1$\sigma$; the plateau mass is taken as the effective mass at $z=4$,
since this effective mass has the smallest uncertainty among the three masses belonging to the plateau. 

\begin{figure}[tb]
\centering
\includegraphics[width=7.5cm,bb=40 40 700 620]{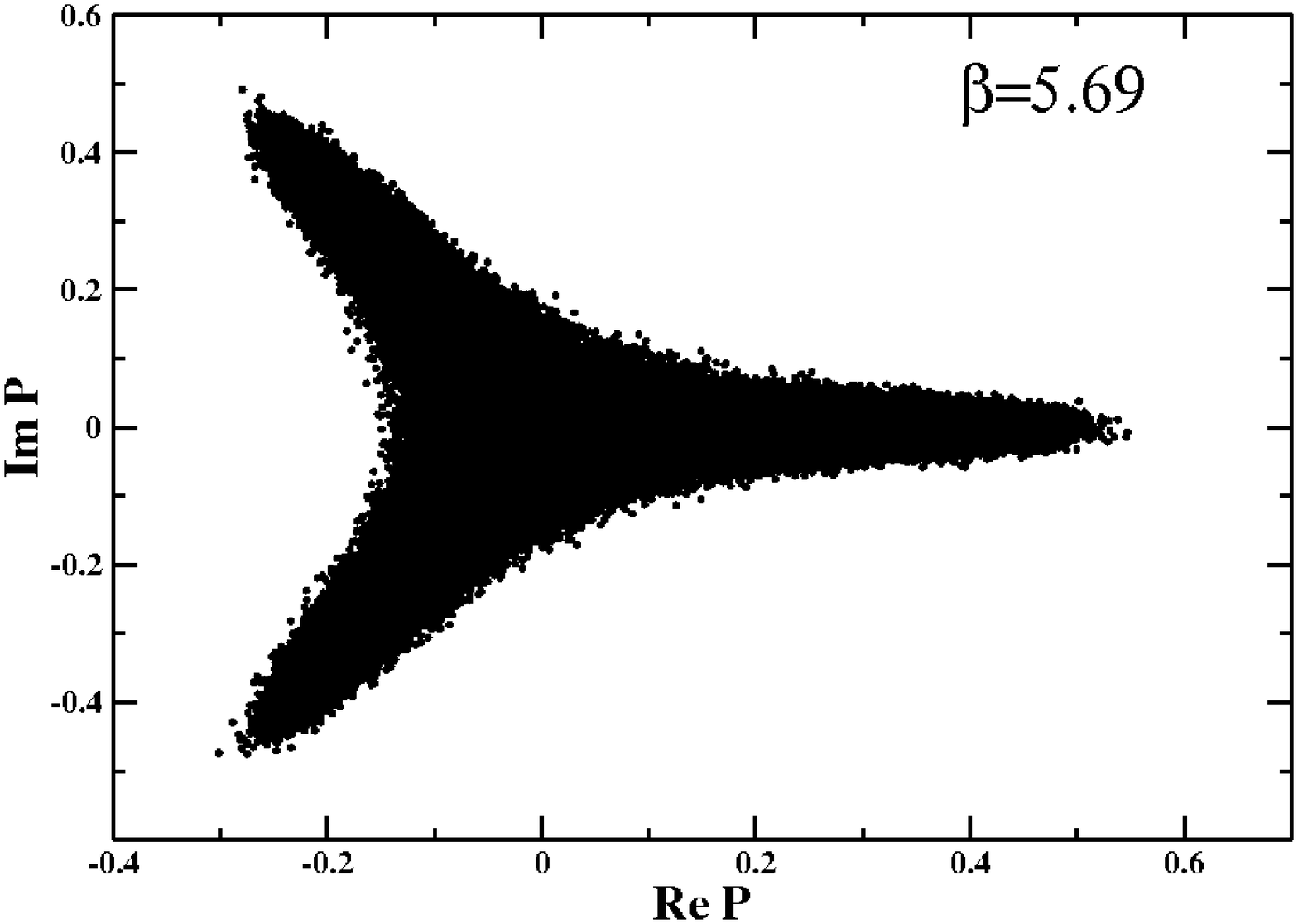} (a)

\includegraphics[width=7.5cm,bb=40 40 700 620]{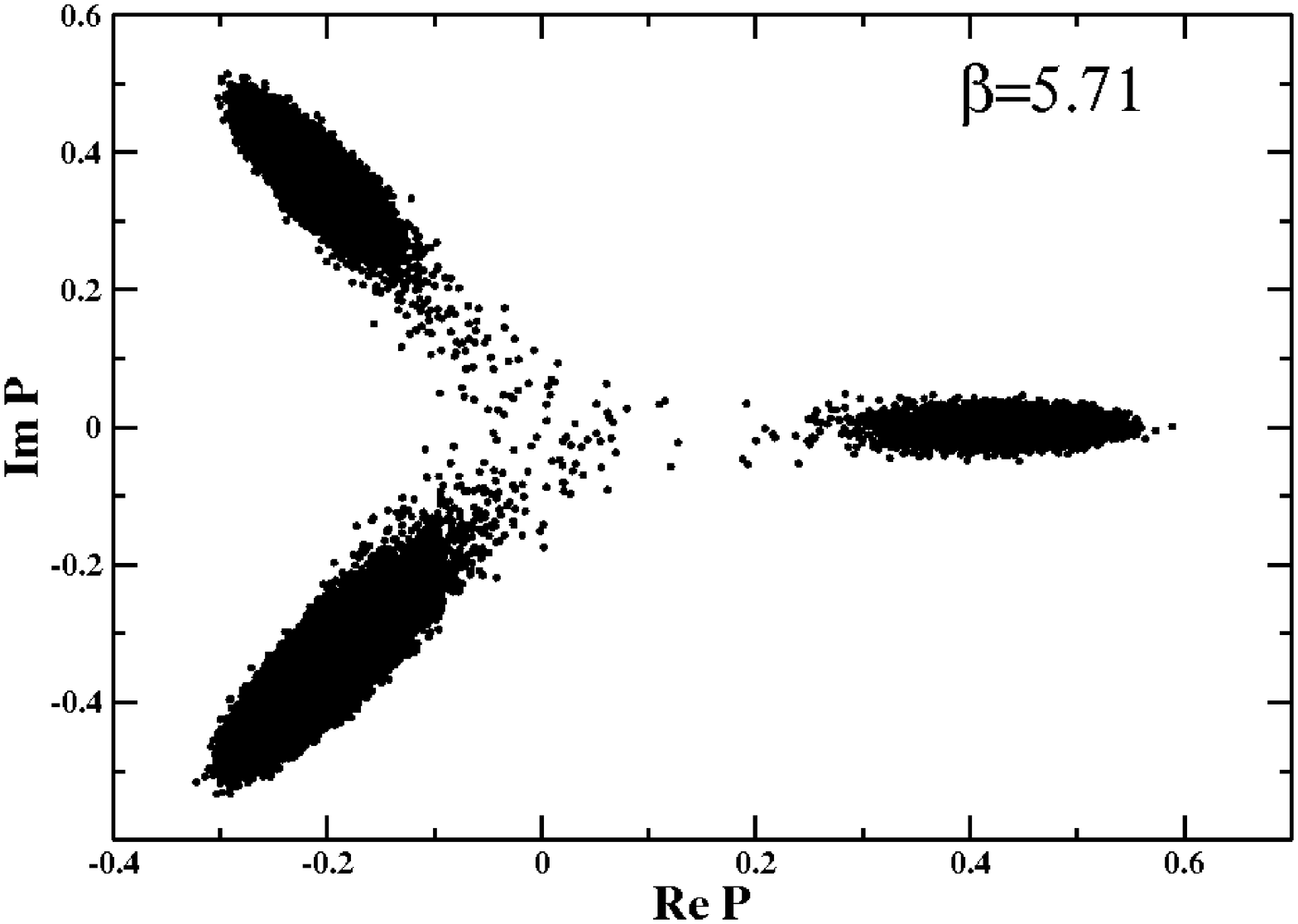} (b)

\includegraphics[width=7.5cm,bb=40 40 700 620]{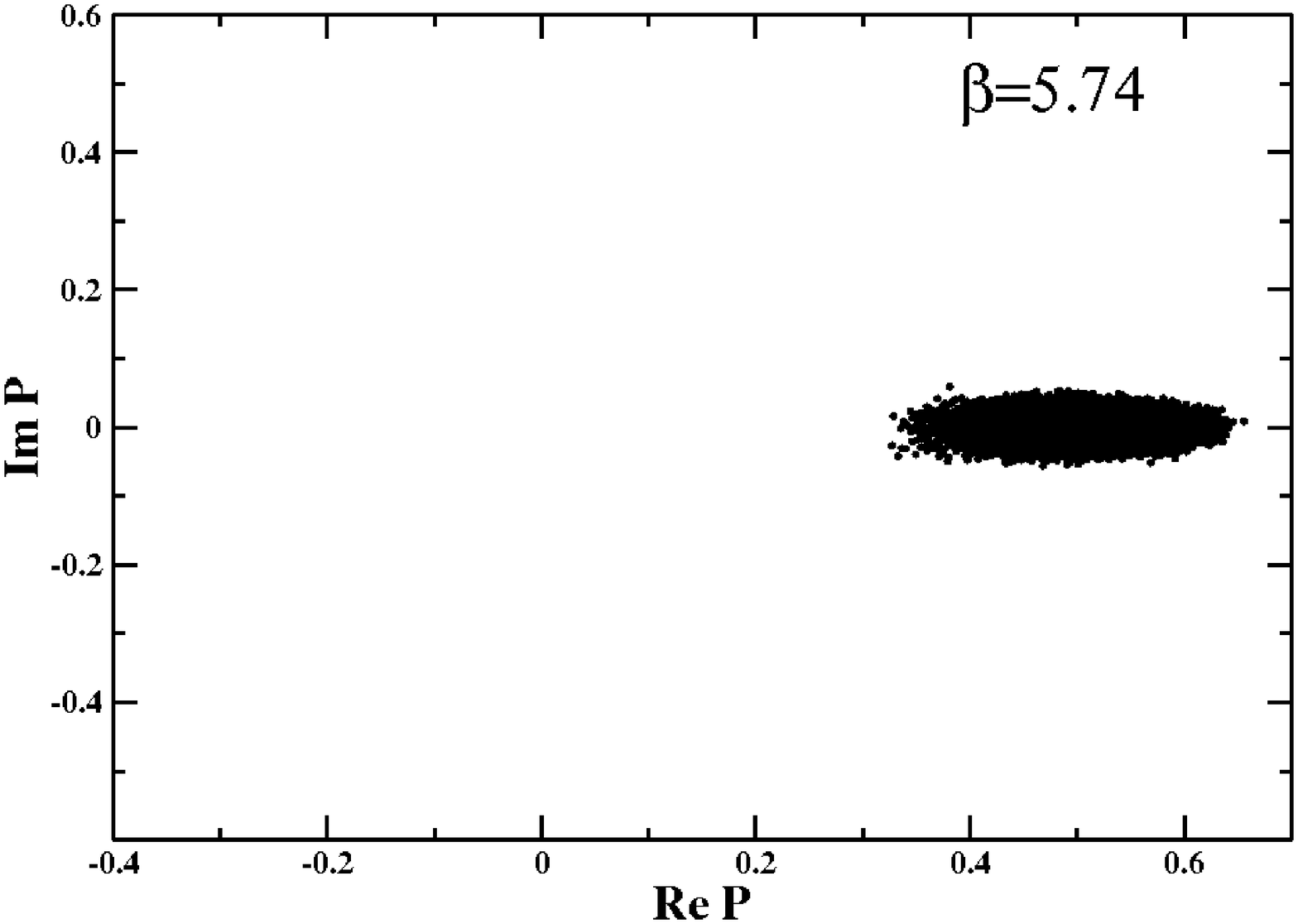} (c)
\caption[]{Scatter plots of the real and imaginary part of the Polyakov loop at 
(a) $\beta=5.69\simeq\beta_t$, (b) $\beta=5.71$ and (c) $\beta=5.74$.}
\label{scatter_plots} 
\end{figure}

Just above the critical value $\beta_t$ we find a large correlation length, which is not of 
physical relevance. It is instead a genuine finite size effect~\cite{Gavai-Karsch-Petersson} 
related to {\it tunneling} between degenerate vacua. This effect disappears by going to larger 
lattice volumes or moving away from $\beta_t$ in the deconfined phase. 
Indeed, by increasing $\beta$ one can see that 
the symmetric phase becomes less and less important in the Monte Carlo ensemble, up to 
disappearance; then, for large enough $\beta$, also the structure with three-degenerate Z(3) 
sectors disappears, leaving only the real sector of Z(3). This is illustrated in 
Fig.~\ref{scatter_plots} where the scatter plot of the Polyakov loop is shown for three 
representative $\beta$ values.

\begin{figure}[tb]
\centering
\includegraphics[width=7.5cm,bb=40 40 700 620]{scatter_5.71.eps} (a) 

\includegraphics[width=7.5cm,bb=40 40 700 620]{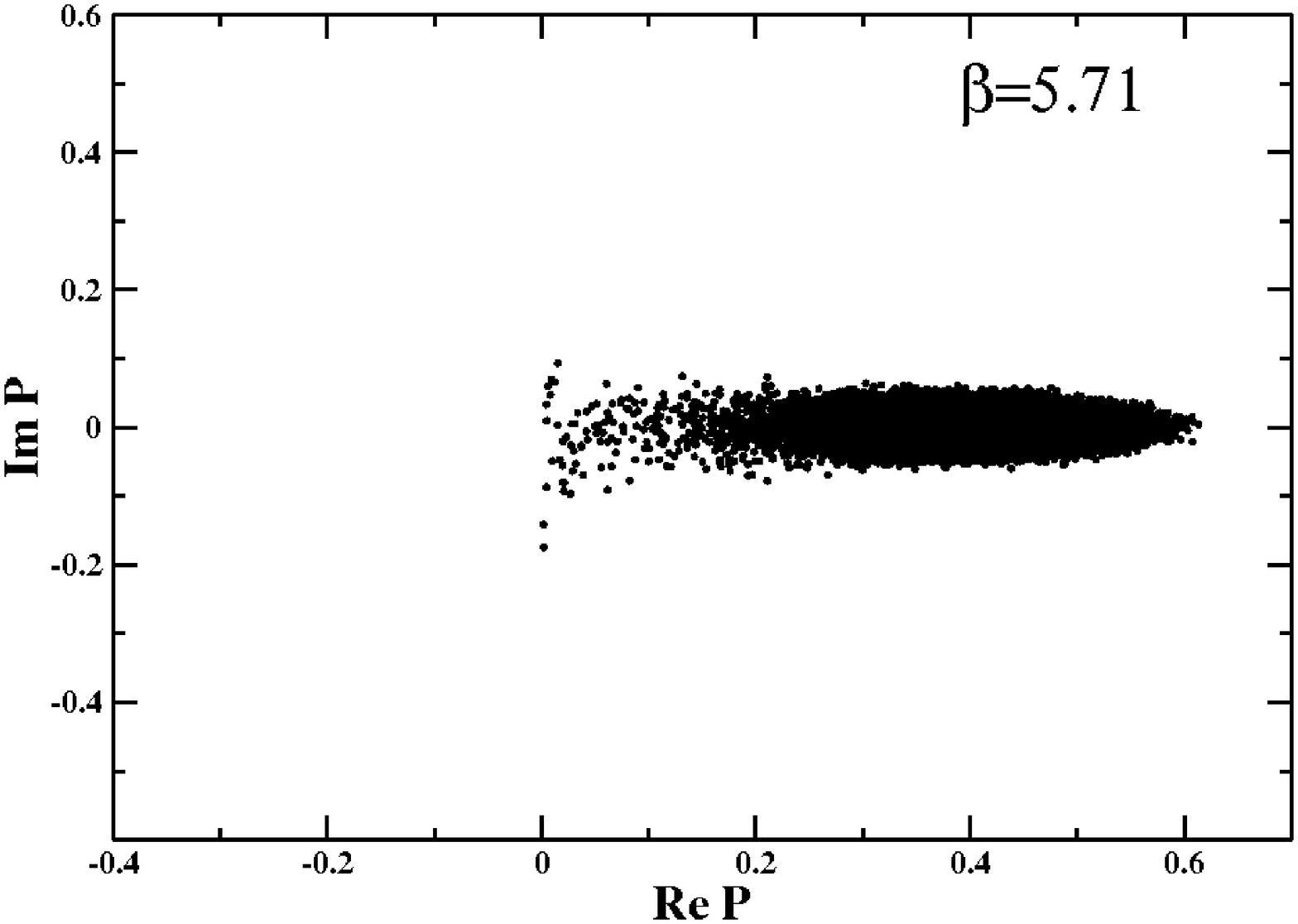} (b)

\caption[]{(a) Typical scatter plot of the complex order parameter $P$ for
$\beta$ larger than 5.705 on a 16$^3\times 4$ lattice. 
There are no states in the symmetric phase,
but tunneling survives between the three broken minima.\newline
(b) Same as (a) with the tunneling between broken minima removed by 
the ``rotation'' to the real phase.}
\label{scatter_plot_rotated} 
\end{figure}

Tunneling can occur between the symmetric and the broken phase, and between
the three degenerate vacua of the deconfined phase. The former survives only in a short range
of temperatures around $\beta_t$, and at $\beta \simeq 5.71$ has almost completely disappeared,
as we can see from Fig~\ref{scatter_plots}(b). When tunneling is active,
the correlation function has the following expression~\cite{Gavai-Karsch-Petersson}:
\begin{equation}
G(|z_1-z_2|)\sim a_0e^{-m_0 |z_1-z_2|}+b_0 e^{-m_t |z_1-z_2|}\;,
\label{corr_funct_tunneling}
\end{equation}
where $m_t$ is the inverse of the tunneling correlation length and is generally much smaller 
than the fundamental mass $m_0$ and therefore behaves as a constant additive term
in the correlation function.
In (\ref{corr_funct_tunneling}) we have taken into account only the 
lowest masses in the spectrum and omitted the ``echo'' terms.
The dependence on $m_t$ in the correlation function can be removed by extracting the
effective mass by use of the combination
\begin{equation}
m_{\mbox{\footnotesize eff}}(z)= \ln \frac{G(z)-G(z+1)}{G(z+1)-G(z+2)} \ .
\label{m_eff_1}
\end{equation}
For $\beta \gtrsim 5.71$ the only active tunneling is among the three broken minima. However,
the separation among them is so clear (see, for instance, Fig.~\ref{scatter_plot_rotated}(a)), 
that it is possible to ``rotate'' unambiguously all the configurations to the real sector 
(see Fig.~\ref{scatter_plot_rotated}(b)) and to treat them on the same ground.

We have performed simulations for several values of $\beta$ in the deconfined phase, up to
$\beta=9.0$, with a statistics of a few hundred thousand ``measurements'' at each $\beta$
value. Simulations have been performed on the PC cluster ``Majorana'' of the INFN Group of
Cosenza. 

\begin{figure}[tb]
\centering
\includegraphics[width=11.5cm,bb=40 40 700 620]{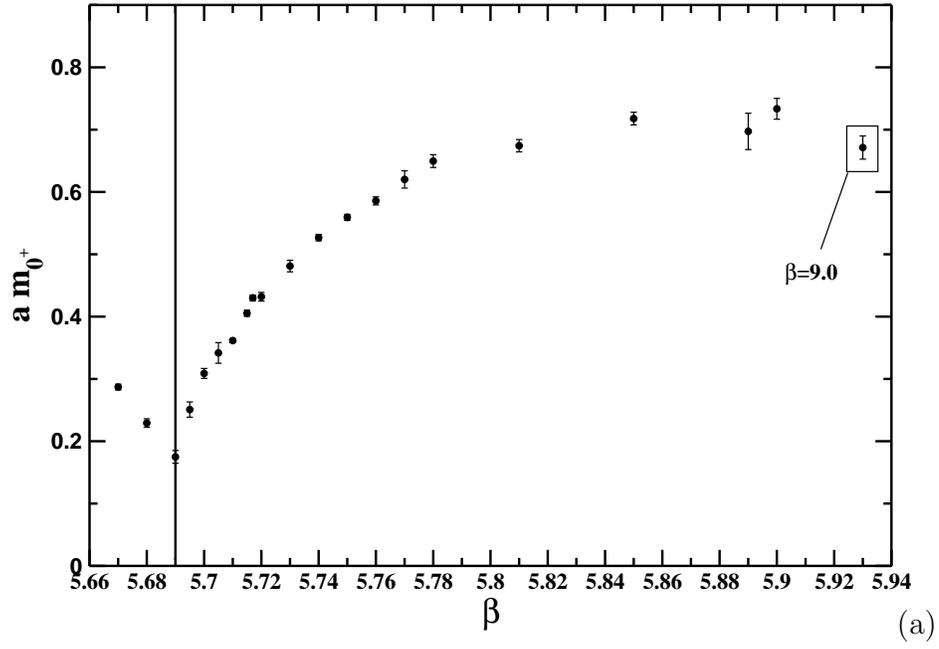} (a)
\includegraphics[width=11.5cm,bb=40 40 700 620]{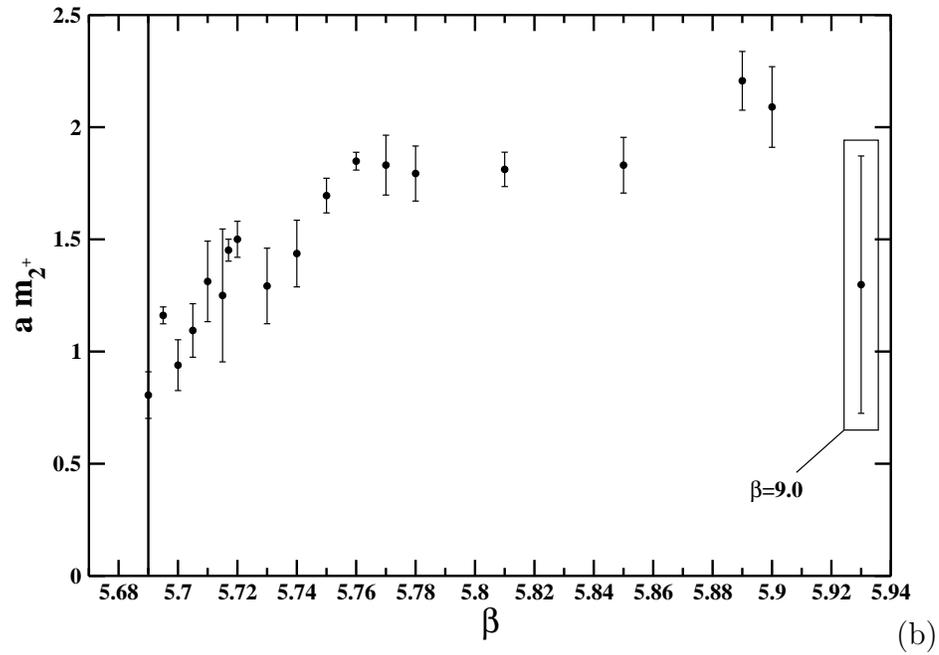} (b)
\caption[]{Screening mass in the 0$^+$ (a) and in the 2$^+$ (b) channel as a 
function of $\beta$.}
\label{masses_vs_beta} 
\end{figure}

\begin{figure}[tb]
\centering
\includegraphics[width=11.5cm,bb=40 40 700 620]{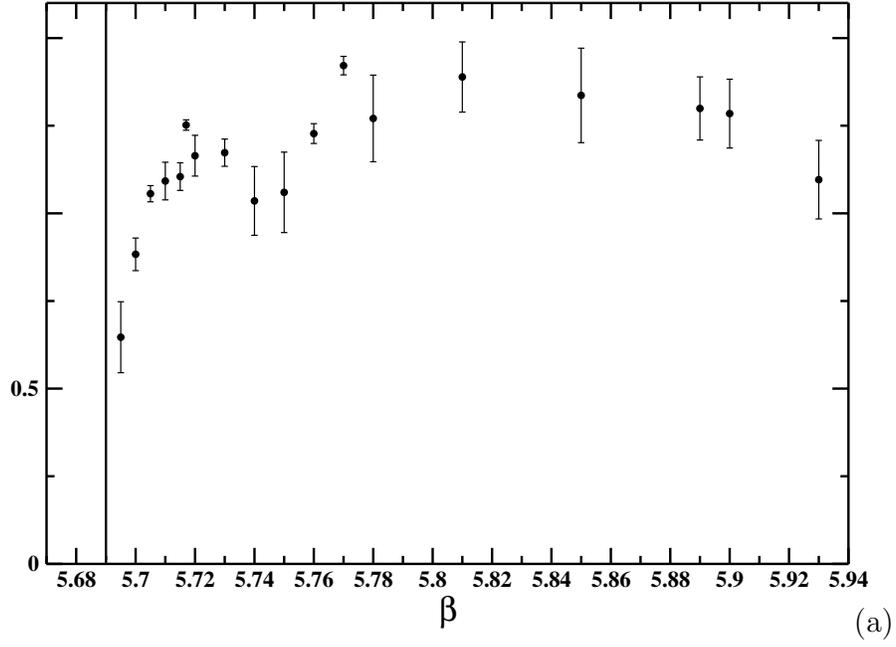} (a)
\includegraphics[width=11.5cm,bb=40 40 700 620]{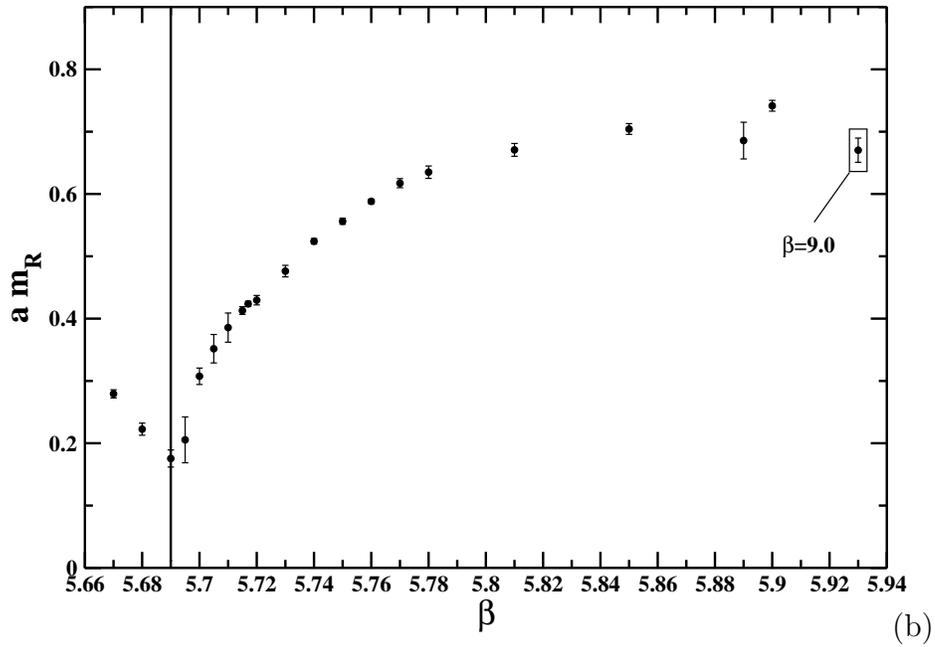} (b)
\caption[]{Screening mass $\hat m_I$ (a) and $\hat m_R$ (b) as a function of $\beta$.}
\label{masses_ri_vs_beta} 
\end{figure}

As discussed in the Introduction and in Section~\ref{sec:masses}, our typical observable is
a screening mass in a given channel of angular momentum and parity, identified as the plateau 
value in the plot of the effective mass as a function of the separation $z$ between two-dimensional
walls. The procedure to determine the plateau value has been illustrated in 
Section~\ref{sec:masses}. A typical example of the behavior of the effective mass with $z$
is shown in Fig.~\ref{eff_masses_5.74} for the 0$^+$ and the 2$^+$ channels
and in Fig.~\ref{eff_masses_ri_5.74} for the masses $\hat m_I$ and $\hat m_R$ extracted from
Eqs.~(\ref{Dumitru2:1}) and (\ref{Dumitru2:2}).
The deviation from the plateau value at small $z$ can be 
attributed to lattice artifacts and to the possible effect of other states with excited
masses in the same channel. Our determinations refer only to the fundamental masses
in a given channel. The behavior with $\beta$ of the fundamental masses in the 0$^+$ 
and the 2$^+$ channels is given in Fig.~\ref{masses_vs_beta}. In Fig.~\ref{masses_ri_vs_beta}
we show instead the masses $\hat m_R$ and $\hat m_I$ for varying $\beta$. 
A summary of all our results is presented in Table~\ref{masses}. We observe from Table~\ref{masses}
that $\hat m_{0^+}$ and $\hat m_R$ are consistent within statistical errors, this indicating 
that the Polyakov loop correlation is dominated by the correlation between the real parts.
Our results for $\hat m_{0^+}$ at $\beta=5.70$ and $\beta=5.90$ ($T\simeq 3T_t/2$) can be compared 
with the corresponding determinations on a $4 \times 8^2 \times 16$ lattice of Ref.~\cite{Datta:1998eb}, 
which give 0.27$^{+0.01}_{-0.02}$ and 0.64(1), respectively. The $\sim$15$\%$ disagreement can be 
taken as a measure of the systematic effects involved in the whole procedure for the 
determination of masses. We cannot make a direct comparison with the determination for $\hat m_{0^+}$
at $\beta=5.93$ on a 16$^3\times 4$ lattice of Ref.~\cite{Grossman:1993wm}, since we did not perform 
simulations at this value of $\beta$. We observe, however, that the determination of 
Ref.~\cite{Grossman:1993wm}, which gives 0.73(5) according to our definition of plateau
mass value, agrees with the results for the two adjacent $\beta$ values in our Table, 
$\beta=5.90$ and $\beta=9.0$.

We can see that the fundamental mass in the 0$^+$ channel, as well as $\hat m_R$, 
becomes much smaller than 1 at $\beta_t$, as expected for a weakly 
first order phase transition. 
In the cases of $\hat m_{0^+}$ and of $\hat m_R$ we have made some determinations 
{\it below} $\beta_t$ (see Figs.~\ref{masses_vs_beta} and \ref{masses_ri_vs_beta}). 
It turns out that masses in lattice units take their minimum value just at $\beta_t$, where 
there is a ``cusp'' in the $\beta$-dependence. Such a behaviour was observed also by the authors
of Ref.~\cite{Datta:2002je}, whose results, when the comparison is possible, agree with ours. 

\begin{table}[htb]
\centering
\caption[]{Summary of the results for the fundamental masses in the $0^+$ and $2^+$ channels, 
for $\hat m_I$ and $\hat m_R$ and for some mass ratios.}

\vskip 0.4cm

\begin{tabular}{|l|l|l|l|l|l|l|r|}
\hline
$\beta$ & $\hat m_{0^+}$  & $\hat m_{2^+}$ & $m_{2^+}$/$m_{0^+}$ & $\hat m_I$ & $\hat m_R$ 
& $m_I/m_R$ & stat. \\
\hline
5.67  & 0.2870(46) &  -        &    -     &    -      & 0.2793(65) &     -     & 750k \\
5.68  & 0.2293(67) &  -        &    -     &    -      & 0.2226(97) &     -     & 250k \\
5.69  & 0.175(10)  & 0.81(10)  & 4.61(86) &    -      & 0.176(14)  &     -     & 770k \\
5.695 & 0.251(12)  & 1.161(38) & 4.63(38) & 0.65(10)  & 0.205(37)  & 3.07(87)  & 320k \\
5.70  & 0.3088(80) & 0.94(11)  & 3.04(45) & 0.883(47) & 0.307(13)  & 2.87(27)  & 770k \\
5.705 & 0.342(17)  & 1.09(12)  & 3.20(50) & 1.056(23) & 0.352(23)  & 2.86(15)  & 750k \\
5.71  & 0.3615(30) & 1.31(18)  & 3.63(53) & 1.092(54) & 0.386(23)  & 2.98(19)  & 770k \\
5.715 & 0.4054(51) & 1.25(30)  & 3.08(77) & 1.105(40) & 0.4129(61) & 2.68(14)  & 250k \\
5.717 & 0.4301(46) & 1.452(49) & 3.38(15) & 1.252(15) & 0.4237(43) & 2.955(65) & 320k \\
5.72  & 0.4319(69) & 1.500(80) & 3.47(24) & 1.164(58) & 0.4296(76) & 2.71(18)  & 770k \\
5.73  & 0.4811(92) & 1.29(17)  & 2.69(40) & 1.173(39) & 0.4762(92) & 2.46(13)  & 170k \\
5.74  & 0.5267(52) & 1.44(15)  & 2.73(31) & 1.035(98) & 0.5241(44) & 1.98(20)  & 770k \\
5.75  & 0.5592(50) & 1.695(77) & 3.03(17) & 1.06(11)  & 0.5562(49) & 1.91(22)  & 770k \\
5.76  & 0.5857(65) & 1.849(39) & 3.16(10) & 1.227(28) & 0.5880(36) & 2.087(61) & 770k \\
5.77  & 0.620(14)  & 1.83(13)  & 2.95(28) & 1.421(26) & 0.6173(73) & 2.303(70) &  60k \\ 
5.78  & 0.650(10)  & 1.79(12)  & 2.76(23) & 1.27(12)  & 0.635(10)  & 2.00(23)  &  60k \\ 
5.81  & 0.6742(99) & 1.812(77) & 2.69(15) & 1.39(10)  & 0.671(10)  & 2.07(18)  & 190k \\ 
5.85  & 0.718(10)  & 1.83(12)  & 2.55(21) & 1.34(13)  & 0.7042(87) & 1.90(21)  & 110k \\ 
5.89  & 0.697(29)  & 2.21(13)  & 3.17(32) & 1.299(90) & 0.686(29)  & 1.89(21)  & 190k \\ 
5.90  & 0.734(17)  & 2.09(18)  & 2.85(31) & 1.284(98) & 0.7415(87) & 1.73(15)  & 215k \\ 
9.0   & 0.671(19)  & 1.30(57)  & 1.93(91) & 1.10(11)  & 0.670(19)  & 1.64(23)  & 200k \\ 
\hline
\end{tabular}
\label{masses}
\end{table} 

\subsection{Scaling behavior and comparison with the Potts model}

It would be interesting to find a scaling law for the fundamental mass in the $0^+$ channel, 
which looks like the one which holds for a second order phase transition, with a
suitable critical exponent. There is, however, an important caveat: while the correlation length 
in lattice units diverges at a second order critical point, it keeps finite at a first order 
transition point. Therefore, any second-order-like scaling law, when applied to the region near 
a first order phase transition, should be taken as an {\it effective} description, which cannot 
hold too close to the transition point.

\begin{figure}[tb]
\centering
\includegraphics[width=15cm, bb=40 40 700 620]{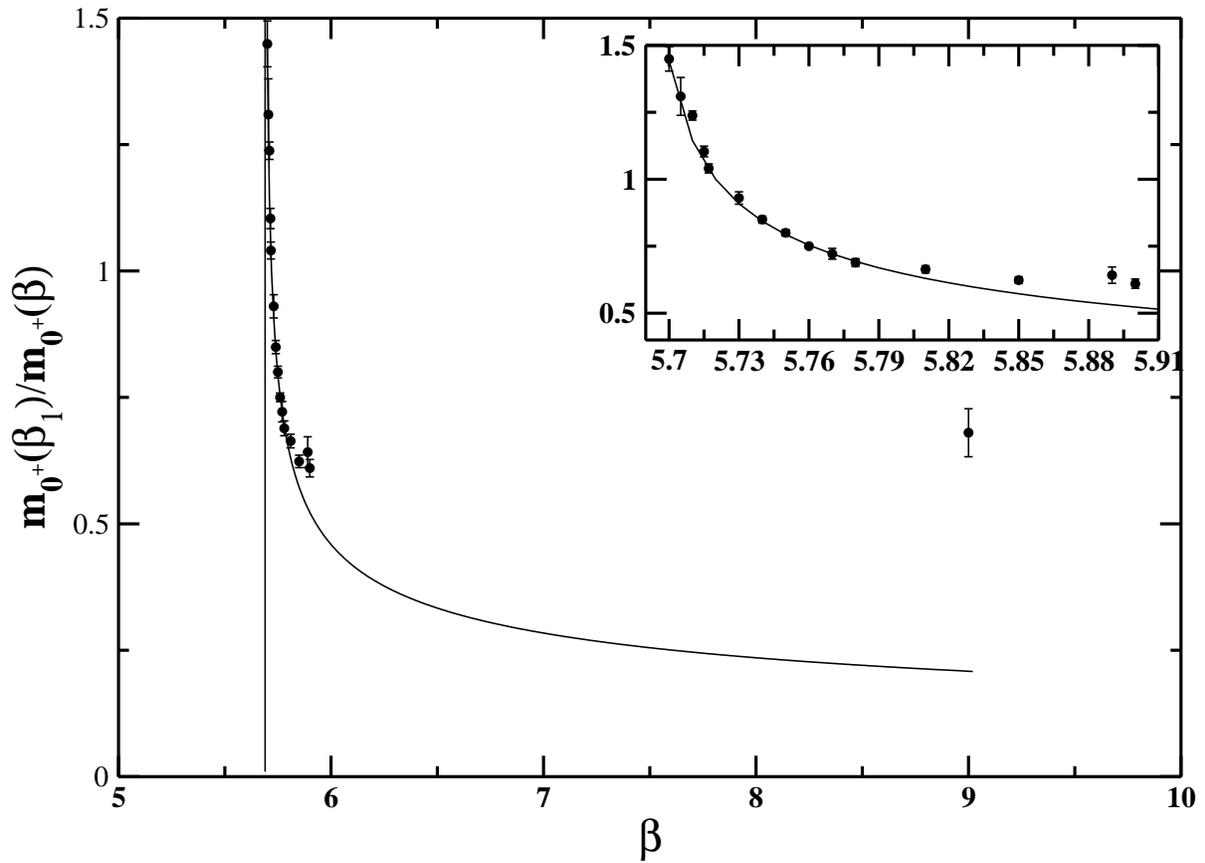}
\caption[]{Comparison between the scaling function $[(\beta_1-\beta_t)/(\beta-\beta_t)]^{1/3}$
and the mass ratio $m_{0^+}(\beta_1)/m_{0^+}(\beta)$ for varying $\beta$, with
$\beta_1$=5.72.}
\label{scaling} 
\end{figure}

With this spirit, we have compared our data with the scaling law
\begin{equation}
\label{scal_rel}
\Bigg( \frac{\beta_1-\beta_t}{\beta_2-\beta_t} \Bigg)^{\nu} \sim 
\frac{\hat m_{0^+}(\beta_1)}{\hat m_{0^+}(\beta_2)} \ ,
\end{equation} 
where $\hat m_{0^+}(\beta_1)$ and $\hat m_{0^+}(\beta_2)$ are the 
fundamental masses in the $0^+$ channel at $\beta_1$ and $\beta_2$, respectively. 
This scaling law~(\ref{scal_rel}) is an approximation of the law
\begin{equation}
\label{scal_rel_corr}
\Bigg( \frac{T_1-T_t}{T_2-T_t} \Bigg)^{\nu} \sim 
\frac{\hat m_{0^+}(T_1)}{\hat m_{0^+}(T_2)} \ ,
\end{equation} 
valid in an interval around $T_t$ short enough that the linear approximation 
$T_1-T_t\propto\beta_1-\beta_t$ holds. We have considered several choices of $\beta_1$
and found that for each of them there is a wide ``window'' of $\beta$ values above $\beta_t$ where
the scaling law~(\ref{scal_rel}) works, with a ``dynamical'' exponent $\nu$. Our results
are summarized in Table~\ref{scaling_fits}. For $\beta_1$=5.72 and 5.73, which lie well inside
this ``window'', the parameter $\nu$ is compatible with $\nu$=1/3, suggested in Ref.~\cite{Fisher-Berker} 
to apply to the {\it standard} correlation function. For $\beta_1$=5.72 we have calculated also 
the $\chi^2$/d.o.f. when $\nu$ is put exactly equal to 1/3, getting $\chi^2$/d.o.f.=0.75 in the window 
from $\beta=5.715$ to $\beta=5.78$. In Fig.~\ref{scaling} we show, for this choice of $\beta_1$, 
the comparison between data and the ``scaling'' function with $\nu$ set equal to 1/3. 

\begin{table}[htb]
\centering
\caption[]{Summary of the fits of the mass ratios ${\hat m_{0^+}(\beta_1)}/{\hat m_{0^+}(\beta)}$ 
with the function $((\beta_1-\beta_t)/(\beta-\beta_t))^\nu$. The second column
contains the largest window of $\beta$ values for which the fit has a $\chi^2$/d.o.f. lower
than 1.}

\vskip 0.4cm

\begin{tabular}{|l|c|l|l|}
\hline
$\beta_1$ & window of $\beta$ values & $\nu$ & $\chi^2$/d.o.f. \\
\hline
5.75  & 5.70  - 5.78 &  0.3619(90) & 0.90 \\
5.74  & 5.70  - 5.78 &  0.365(11)  & 0.92 \\
5.73  & 5.695 - 5.85 &  0.329(12)  & 0.77 \\
5.72  & 5.695 - 5.78 &  0.354(12)  & 0.73 \\
5.717 & 5.715 - 5.81 &  0.3228(94) & 0.69 \\
5.715 & 5.695 - 5.78 &  0.358(10)  & 0.86 \\
5.71  & 5.72  - 5.78 &  0.3951(76) & 0.33 \\
5.705 & 5.705 - 5.78 &  0.448(22)  & 0.89 \\  
5.70  & 5.695 - 5.90 &  0.3141(58) & 0.94 \\
5.695 & 5.695 - 5.90 &  0.3095(67) & 0.41 \\
\hline
\end{tabular}
\label{scaling_fits}
\end{table} 

\begin{figure}[tb]
\centering
\includegraphics[width=16cm, bb=40 40 700 620]{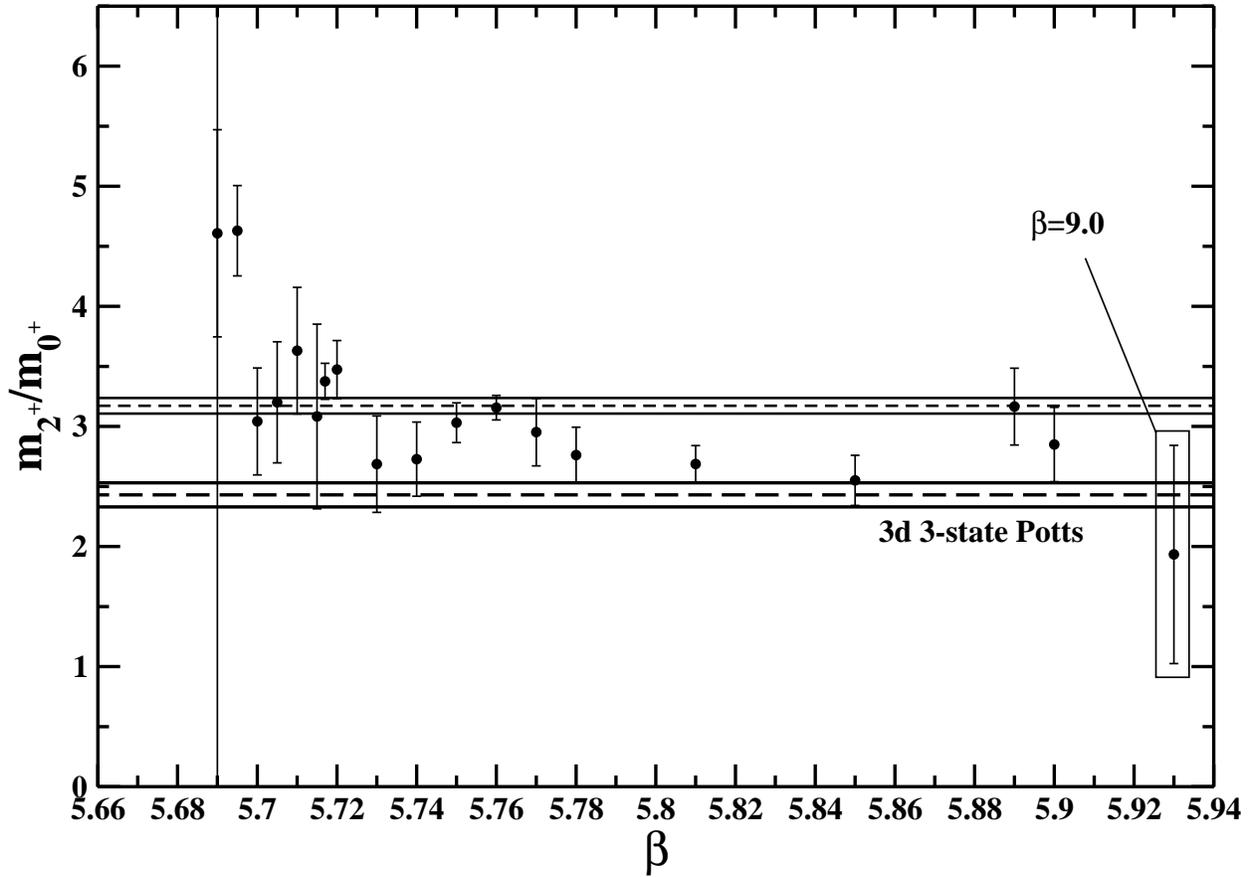}
\caption[]{Ratio $m_2^+/m_0^+$ as a function of $\beta$ in the deconfined phase. The three upper
horizontal lines represent the constant (with its error) which fits the data (see the text for details);
the three lower horizontal lines represent the corresponding mass ratio (with its error) found in the 3$d$ 
3-state Potts model~\cite{FFGP06}.}
\label{ratio_20} 
\end{figure}

Then, we have considered the $\beta$-dependence of the ratio $m_{2^+}/m_{0^+}$, shown in
Fig.~\ref{ratio_20}. We have found that this ratio can be interpolated with a constant
in the interval from $\beta_t$ to $\beta=5.77$. This constant turned out to be
3.172(65), with a $\chi^2$/d.o.f equal to 1.085. In the fit we excluded the point at $\beta$=5.695, 
for which the determination of $m_{2^+}$ is probably to be rejected (see Fig.~\ref{masses_vs_beta}(b)).  
If the point at $\beta$=5.695 is included, the constant becomes 3.214(64) with $\chi^2$/d.o.f =2.21.
The fact that the ratio $m_{2^+}/m_{0^+}$ is compatible with a constant in the mentioned interval
suggests that $\hat m_{2^+}$ scales similarly to $\hat m_{0^+}$ near the transition. This constant
turns out to be larger than the ratio between the lowest massive excitations in the
same channels in the broken phase of the 3$d$ 3-state Potts model, which was determined
in Ref.~\cite{FFGP06} to be 2.43(10).

\subsection{Comparison with Polyakov loop models}

\begin{figure}[tb]
\centering
\includegraphics[width=16cm, bb=40 40 700 620]{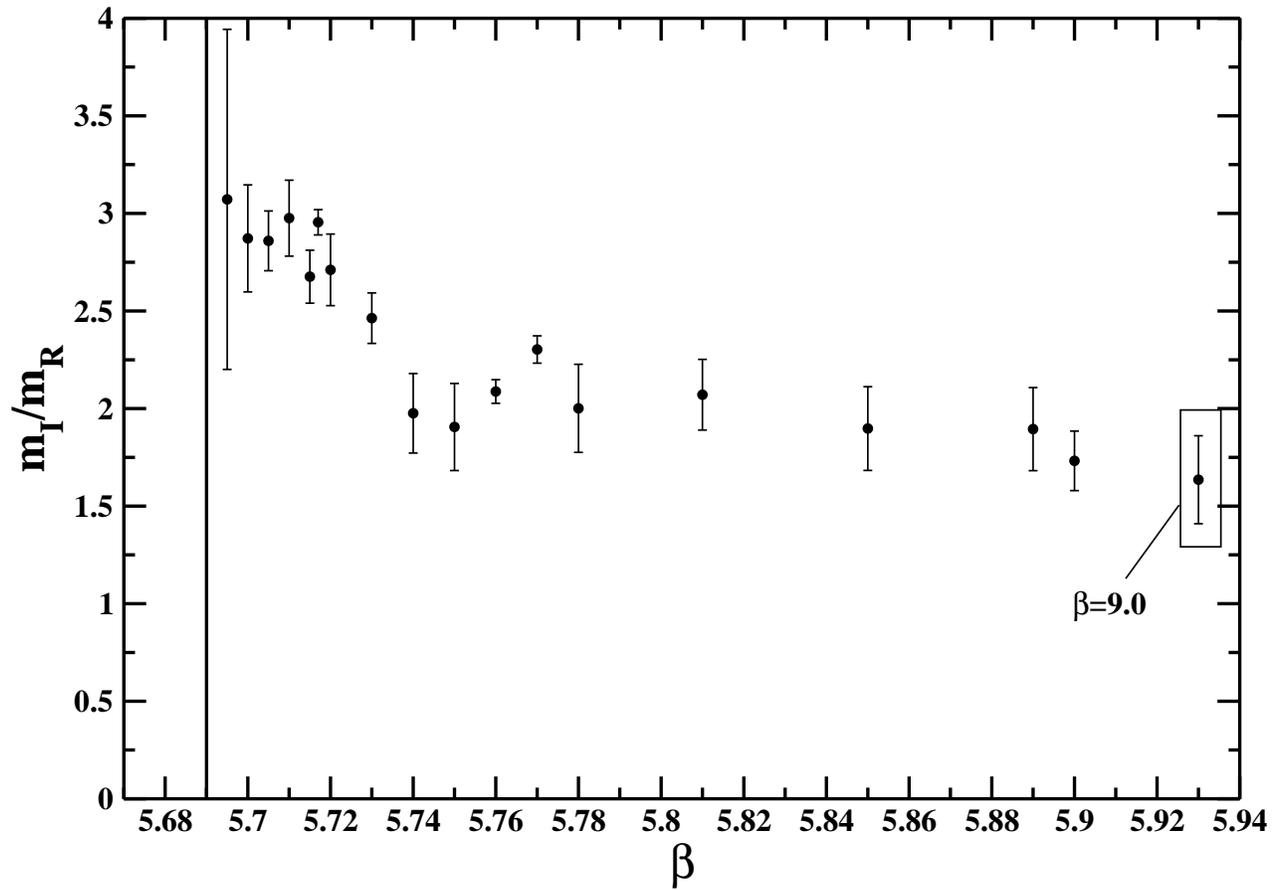}
\caption[]{Ratio $m_I/m_R$ as a function of $\beta$ in the deconfined
phase. The vertical line corresponds to the critical $\beta$ value.}
\label{ratio_ri} 
\end{figure}

We have calculated the ratio $m_I/m_R$ for $\beta$ ranging from 5.695 up to 9.0. We observe from 
Fig.~\ref{ratio_ri} that this ratio is compatible with 3/2 at the largest $\beta$ values considered, 
in agreement with the high-temperature perturbation theory. Then, when the temperature is lowered 
towards the transition, this ratio goes up to a value compatible with 3, in agreement with the 
Polyakov loop model of Ref.~\cite{Pisarski0110214}, which contains only quadratic, cubic and 
quartic powers of the Polyakov loop, i.e. the minimum number of terms required in order to be 
compatible with a first order phase transition. The same trend has been observed also in 
Ref.~\cite{Datta:2002je}.

\section{Conclusions and outlook}

In this work we have studied some screening masses of the (3+1)$d$ SU(3) pure gauge theory
from Polyakov loop correlators in a large interval of temperatures above the deconfinement 
transition. 

In particular, we have considered the lowest masses in the 0$^+$ and the 2$^+$
channels of angular momentum and parity and the screening masses resulting from the
correlation between the real parts and between the imaginary parts of the Polyakov loop.

The behavior of the ratio between the masses in the 0$^+$ and the 2$^+$ channels with the 
temperature suggests that they have a common scaling above the transition temperature. 
This ratio turns to be $\simeq$30\% larger than the ratio of the lowest massive excitations in 
the same channels of the 3$d$ 3-state Potts model in the broken phase. This can be taken
as an estimate of the level of approximation by which the Svetitsky-Yaffe conjecture, valid in strict 
sense only for continuous phase transitions, can play some role also for (3+1)$d$ SU(3) at finite 
temperature. 

The dependence on the temperature of the ratio between the screening masses from the
correlation between the real parts and between the imaginary parts of the Polyakov loop 
shows a nice interplay between the high-temperature regime, where perturbation theory should work, 
and the transition regime, where mean-field effective Polyakov loop models could 
apply.

\end{document}